\def\b #1{{\rm\bf #1 }}
\def\x{{\vec x}}
\def\la{{\langle }}
\def\ra{{\rangle }}

\def\e{{\hat e  }}
\def\tA{{\tilde A}}
\def\be{\begin{equation}}
\def\ee{\end{equation}}
\def\bea{\begin{eqnarray}}
\def\eea{\end{eqnarray}}
\def\det{\Delta}

\documentstyle[aps,prl,12pt,epsfig]{revtex}
\begin{document}
\draft
\title{On the Anti-Wishart distribution}
\author{Yi-Kuo Yu$^1$ and Yi-Cheng Zhang$^2$}
\address{ $^1$Department of Physics, Florida Atlantic University,
Boca Raton, FL 33431, USA\\
$^2$Institut de Physique Th\'eorique, Universit\'e de Fribourg,
CH-1700, Fribourg, Switzerland}
\maketitle

\vspace*{0.5in}

\begin{abstract}
We provide the probability distribution function of matrix elements
 each of which is the inner product of two vectors.
 The vectors we are considering here are
 independently distributed but not necessarily Gaussian variables.
 When the number of components $M$ of each vector is greater than
 the number of vectors $N$, one has a $N\times N$ symmetric matrix.
 When $M\ge N$ and  the components of each vector are independent
 Gaussian variables,  the distribution function of the $N(N+1)/2$
 matrix elements  was obtained by Wishart in 1928.
 When $N>M$, what we called the ``Anti-Wishart'' case,
 the matrix elements are no longer completely
 independent because the true degrees of freedom  becomes smaller
 than the number of matrix elements.
 Due to this singular nature, analytical derivation of the probability
 distribution function is much more involved than the corresponding
  Wishart case. For a class of general random vectors,
 we  obtain the analytical distribution function in a closed form,
which is  a product of various
factors and delta function constraints, composed of various determinants.
 The distribution function of the matrix element for the
$M\ge N$ case with the same class of random vectors
 is also obtained as a by-product.
 Our result is closely related to
and should be valuable for the study of random magnet problem and
 information redundancy problem.
\end{abstract}

\section{Introduction}
Many problems in physics~\cite{phys}
  can be related to the matrix problem that we
 will discuss. The matrix we shall consider in this work takes a special form:
 each  matrix element, $Y_{i,j}$ is
 the inner product of two independent random vectors.
  Historically speaking, this type of random
 matrix came into research literature even before the
 now-called random matrix theory first introduced and
explored by Wigner, Dyson, Mehta and others~\cite{Mehta}.
 The number of degrees of freedom
is important for this class of matrices. Denote $N$ the dimension of
 square matrix $\b Y$,
 $M$ the dimension of the vectors. Depending which is larger,
 the resultant matrices can be singular with lots of zero eigenvalues;
 or normally behaved.

The original motivation to study this type of distribution function
mainly come from the effort to understand the correlations of fluctuations.
But in recent years, interests have manifested in studying the case where
matrix elements are not independent, but inner products of vector-pairs.
To be more precise, let us consider the following precise definition.
Mathematically, we may denote  N vectors $\x_1,\x_2,\cdots,\x_N$,
each of which lives
in $M$ dimensions.  For example, vector $\x_i$ has its components
$x_i^1,x_i^2,\cdots, x_i^M$. After shifting by the average in each sample,
 and possible rescaling,
we may  assume that the vector components
$x_i^\alpha$
(for all $i \in \{1,2,\cdots,N \}$, $\alpha \in \{1,2,\cdots,M\}$)
 are  random variables with some distribution function $P(\{x_i^\alpha\})$.
  We then define the Matrix of interest $\b Y$ whose matrix element $Y_{ij}$
 is simply defined as $\x_i \cdot \x_j$.
\begin{equation}
Y_{ij} \equiv \x_i\cdot \x_j = \sum_{\alpha = 1}^M x_i^\alpha x_j^\alpha.
\end{equation}

When $M>N$ and the background distribution $P(\{x_i^\alpha\})$
 being a  Gaussian
\be \label{background}
P(\{x_i^\alpha\}) = \left[ {{\rm det}^{1/2} [\sigma]  \over (2 \pi )^{N/2}}
\right]^M \exp(-{1\over 2} \sum_{i,j = 1}^N \x_i\cdot \x_j \sigma_{i,j}),
\ee
Wishart obtained~\cite{Wishart} in 1928
 a compact expression for the distribution function
of $P(\{Y_{ij}\})$. The result looks simplest when the matrix $[\sigma]$ is
 an identity matrix, and we have
\be \label{Wishart.dist}
P(\{Y_{ij}\}) = {\cal N}^{-1} \;\;
{\rm det}[{\rm \bf Y}]^{(M-N-1)/ 2} \exp (-{\rm Tr\; {\bf Y}/2})
\ee
 where ${\cal N}$ is some normalization constant.
For the case of general $[\sigma]$, we replace the
$\exp (-{\rm Tr\; {\bf Y}/2})$ part by
$\exp (- \sum_{i,j=1}^N \sigma_{i,j} Y_{i,j}/2 )$.
 A general re-derivation can be found in~\cite{Wilks}.
This result has been the fundamentals (and one of
the triumphs) of multivariate statistical inference.
When the components of the vectors are not Gaussian variables,
 this is still a challenging problem not to mention the
 generic case in the opposite direction $M<N$, which we
 coined as ``Anti-Wishart'' case.

In this paper, we will extend the analytical result to
 a class of more general $P(\{x_i^\alpha \})$ and also to the
``Anti-Wishart'' case. Basically, we will consider the case where
\be \label{bg_general}
P(\{x_i^\alpha \}) = \prod_{i=1}^N f_i(\x_i \cdot \x_i).
\ee
Note that each vector $\x_i$ is allowed to have a different spherical
 distribution function. For example, we allow $f_1$ being a Gaussian,
 $f_2$ being a delta function  ${\cal C}\delta(\x_2\cdot \x_2 - 1)$
 constraining the vector $\x_2$ to have unit length etc.

In Wishart's case, the matrix elements $Y_{ij}$ of ${\bf Y}$ have
enough degrees of freedom. All the eigenvalues are nonzero generically.
This is so because the original degrees of
 freedom, $MN$ is larger than the number of $Y_{ij}$s, $N^2$.
 The case for $M<N$, however, eluded our reach for many years.
 Partly due to the difficulty in dealing with singular measures
 and partly due the lack of motivations.
 However, recent advances in many branches of science have necessitated
 quantitative knowledge about distribution function of this sort.
 One simple example come from the study of bio-molecular interaction
 matrix, e.g., the protein-protein interaction matrix that is
 now intensively studied in molecular biology. The knowledge of such
 matrix is extremely important to quantitatively understand how
 cell function, etc. Another example come from the scenario in global
 knowledge network proposed by Maslov and Zhang~\cite{MZ} where
 information redundancy is exploited. Finally, in the random magnetic system
  the coupling $J_{ij}$ between two spins ${\b S}_i$ and ${\b S}_j$
 could be random variables obtained form inner product of two vectors
 $\x_i$ and $\x_j$ that characterize the property at each specific sites
 $i$ and $j$. This actually happens while transforming a  two dimensional
 $XY$ random field magnetic model into a random bond one\cite{Yu_99}.
In this case, it will be very desirable to have knowledge of such
distribution function.

In this paper, for the class of random
 vector distribution~(\ref{bg_general}),
 we document down {\em for the first time}
the {\em correct} exact distribution of $\{ Y_{ij} \}$ in analytical form
 using only fundamental tools of linear algebra.
 To have the best flow in showing the derivation, we will state
 two useful lemmas and introduce useful notations in the next section
 followed by another section devoted to the derivation. The proofs of the two
 lemma are relegated to the appendix.

\section{Two Lemmas}

Before we get to the derivation of the distribution function,
we would like to introduce some useful notation
and state two useful lemmas.
First, let us denote $\det_{i_1,i_2,\cdots, i_K}$
as the determinant of a compactified matrix obtained by eliminating from
${\bf Y}$ matrix elements whose both indices are not completely in the
set $\{i_1, i_2, \cdots, i_K\}$.
For convenience, we define
$\det_{1,2,\cdots,L-1} = 1$ and $\det_{1,2,\cdots,L-1,j} = Y_{jj}$ when $L=1$.
Naturally the following abbreviations $\det_{1,2,\cdots,L-1} = \det_0$
 and $\det_{1,2,\cdots, L-1,j} = \det_{0,j} = \det_j$ when $L=1$ apply.

The notation $\la i+j\ra$ denotes a
single index
 with the following rules:
$Y_{i\la j+k\ra} \equiv
 Y_{ij}+Y_{ik}$ and $Y_{\la j+k\ra \la j+k\ra} = Y_{jj}+Y_{kj}+
Y_{jk}+Y_{kk}$. Having introduced such notations, we now state
the two  useful lemmas.

\noindent{\bf Lemma 1}
Provided the matrix ${\bf Y}$ is symmetric, using the above definitions we have
\bea
&& 4\; \det_{1,2,\cdots, L-1, k}\;\cdot\;
\det_{1,2,\cdots, L-1,j} - \nonumber \\
&& - \left(\det_{1,2,\cdots, L-1,\la k+j\ra}
-\det_{1,2,\cdots, L-1,k}-\det_{1,2,\cdots,L-1,j}\right)^2 \nonumber \\
&& = 4\;\det_{1,2,\cdots, L-1}\; \cdot \; \det_{1,2,\cdots, L-1,k,j}
\eea

\noindent{\bf Lemma 2}
Using the above definition and assume the matrix ${\bf Y}$ being symmetric,
 we have
\bea
&& \det_{1,2,\cdots, L-1, j}\;\cdot
 \; \left[\det_{1,2,\cdots, L-1,\la k+l\ra}
-\det_{1,2,\cdots, L-1,k}-\det_{1,2,\cdots,L-1,l}\right]
\nonumber \\
&& - {1\over 2} \left[\det_{1,2,\cdots, L-1,\la j+k\ra}
-\det_{1,2,\cdots, L-1,k}-\det_{1,2,\cdots,L-1,j}\right] \cdot \nonumber \\
&& \hspace{0.3in} \cdot \left[\det_{1,2,\cdots, L-1,\la j+l\ra}
-\det_{1,2,\cdots, L-1,k}-\det_{1,2,\cdots,L-1,l}\right] \nonumber \\
&& \; = \det_{1,2,\cdots, L-1}\; \cdot \nonumber \\
&& \;\;\cdot\; \left[\det_{1,2,\cdots, L-1,j, \la k+l \ra}
 -\det_{1,2,\cdots, L-1,j,k} - \det_{1,2,\cdots, L-1, j, l} \right]
\eea

\section{Derivation}
In this section, under the Gaussian background distribution
we will derive the Anti-Wishart distribution and also
obtain the Wishart distribution as a by-product.
Formally, we write the distribution function as
\be
P(\{Y_{ij}\}) = \int d\x_1 d\x_2 \cdots d\x_N \;
 \left[\prod_{i=1}^N f_i(\x_i \cdot \x_i)\right] \; \prod_{i\le j}
 \delta(Y_{ij}-\x_i \cdot \x_j)\label{dist.Yij.def}
\ee
where $\x_i\cdot\x_j = \sum_{\alpha =1}^M x_i^\alpha x_j^\alpha$ is the inner
 product of vector $i$ and vector $j$. When the components of each vector
 are independent Gaussian random variables, one has
 \be
\prod_{i=1}^N f_i(\x_i \cdot \x_i) = (2\pi)^{-MN/2}
 \exp \left[ -{1\over 2}
 \sum_{i=1}^{N} \sum_{\alpha =1}^{M} (x_i^\alpha)^2 \right].
\ee
 We will come back
 to it when compare to the Wishart result.

 Because we want to go beyond the case where components of each vector
 are independent Gaussian random variables,
   our strategy is to integrate one solid angle at a time, see below.
 Since the matrix
elements are invariant under rotation in the $M$ dimensional space, we
can choose $\x_1 \parallel \e_M$ and write the components of the
rest of other $N-1$ vectors in polar angles, i.e., write
\bea
x_i^M &=& r_i\cos \theta_{M-1;\, i}\nonumber \\
x_i^{M-1} &=& r_i \sin \theta_{M-1;\, i} \cos \theta_{M-2;\, i}\nonumber \\
x_i^{M-2} &=& r_i \sin \theta_{M-1;\, i}
\sin \theta_{M-2;\, i} \cos \theta_{M-3;\, i}\nonumber \\
\vdots && \vdots \nonumber \\
x_i^2 &=& r_i \sin \theta_{M-1;\, i}\sin \theta_{M-2;\, i}
\cdots \sin \theta_{2;\, i} \cos \theta_{1;\, i} \nonumber \\
x_i^1 &=& r_i \sin \theta_{M-1;\, i}\sin \theta_{M-2;\, i}
\cdots \sin \theta_{2;\, i} \sin \theta_{1;\, i} \label{coord.in.polar}
\eea
  where the notation
 $\theta_{a;\, i}$ represents the $a$th polar angle of the $i$th
 vector.
 Under such decomposition, the volume element
\be
d\x_i \to r_i^{M-1}d r_i d\Omega_i(M) =  r_i^{M-1}d r_i
\prod_{a=1}^{M-1}[\sin \theta_{a;\, i}^{a-1} d\theta_{a;\, i} ]
\ee
where
$0 \le \theta_{2\le a \le M-1;\, 2\le i \le N} < \pi$ and
$0 \le \theta_{1;\, 2\le i \le N} < 2\pi$ are the polar angles
 and are bounded by the expression above. Naturally, the
 solid angle element $d\Omega_i(b)$ in $b$ dimension is given by
\be
d\Omega_i(b) \equiv \prod_{a=1}^{b-1}[\sin \theta_{a;\, i}^{a-1} d\theta_{a;\, i}].
\ee

One can immediately integrate away first the $N$ radial vectors to get rid
of the delta functions contain only diagonal elements of the matrix. We then
integrate away the solid angles $\theta_{M-1;\, 2\le i \le N}$.
After the first stage, we have
\bea
P(\{Y_{ij}\}) &=& 2^{-N} \left[\prod_{i=1}^N f_i(Y_{ii})\right]
\int
\prod_{i=1}^{N} [Y_{ii}^{M-2\over 2} d\Omega_i(M)]
\prod_{j=2}^{N} \delta\left(Y_{1j}- \sqrt{Y_{11}Y_{jj}}
 \cos \theta_{M-1;\, j} \right) \nonumber \\
&& \cdot \prod_{2\le j <k\le N}
\delta\left(Y_{jk}-\sqrt{Y_{jj}Y_{kk}} I_{j,k}(M-1)
\right) \label{dist.s1}
\eea
where
\be
I_{j,k}(a) = \cos \theta_{a;\, j} \cos \theta_{a;\, k}
     + \sin \theta_{a;\, j} \sin \theta_{a;\, k} I_{j,k}(a-1)
\ee
with
\be
I_{j,k}(a=0) = 1 \hspace*{1in} {\rm for~all~}j,\, k.
\ee

Note that the factor $\prod_{i=1}^N f_i(Y_{ii})$  becomes
$(2\pi)^{-MN/2}e^{-{1\over 2}{\rm Tr}{\b Y}}$ when the components of each
 vector are independent Gaussian variables.
Note that in eq.(\ref{dist.s1}) the delta functions do not depend on the
 polar angles of vector $\x_1$, therefore we can
integrate $d\Omega_1$ and obtain a factor $K_{M}$, area of unit sphere
 in $M$ dimension. Now let us note that we may rewrite
 $Y_{ij} = {1\over 2}[\det_{\la i+j\ra}-\det_{i}-\det_{j}]
= {1\over 2}[\det_{0,\la i+j\ra}-\det_{0,i}-\det_{0,j}] $.
We can therefore rewrite the expression (\ref{dist.s1}) into
\bea
P(\{Y_{ij}\}) &=& 2^{-N} \left[\prod_{i=1}^N f_i(Y_{ii}) \right]
  K_M \left[\prod_{i=1}^N\det_i^{M-2\over 2}\right] \int
\left[\prod_{i=2}^{N} d\Omega_i(M)\right] \nonumber \\
&& \cdot
\prod_{j=2}^{N} \delta\left({1\over 2}[\det_{0,\la 1+j \ra}-\det_{0,1}
-\det_{0,j}] -\sqrt{\det_{0,1} \det_{0,j} }
 \cos \theta_{M-1;\, j}\right) \nonumber \\
&& \cdot \prod_{2\le j <k\le N}
\delta\left({1\over 2}[\det_{0,\la j+k \ra}-\det_{0,j}-\det_{0,k} ]
-\sqrt{\det_{0,j}\det_{0,k}} I_{j,k}(M-1)\right) \label{dist.s2}
\eea

We may then integrate away $\theta_{M-1;\, 2\le j\le N}$
 using
\be
\int_0^\pi \sin \theta_j d\theta_j = {1\over \sqrt{\det_{0,1}\det_{0,j}}}
\int_{-\sqrt{\det_{0,1}\det_{0,j}}}^{\sqrt{\det_{0,1}\det_{0,j}}}
 \;\; d\; y
\ee
 where $y \equiv \sqrt{\det_{0,1}\det_{0,j}} \cos \theta_j $.
Note that because of the delta functions
\bea
 \sqrt{\det_{0,1} \det_{0,j}}\cos \theta_{M-1;\, j} &=&
{1\over 2}[\det_{0,\la 1+j\ra }-\det_{0,1}-\det_{0,j}], \nonumber \\
\sin \theta_{M-1;\, j} &=& \sqrt{1-\cos^2\theta_{M-1;\, j} } =
\sqrt{\det_0 \det_{0,1,j} \over \det_{0,1} \det_{0,j}}
\eea
where   lemma 1 and $0\le \theta_{M-1;\, j} < \pi$
are used in the sine part of the above equation.
Therefore  the integral over polar angles $ \{ \theta_{M-1;\, j} \}_{j=2}^N $
 yields
\bea
P(\{Y_{ij}\})&=&  2^{-N} \left[\prod_{i=1}^N f_i(Y_{ii}) \right]
 K_M \; \det_1^{(N-2)(N-M+1)\over 2}
\left[\prod_{j=2}^N \det_{1,j}^{M-3\over 2}\right]
\int \left[\prod_{i=2}^{N} d\Omega_i(M-1)\right]
\nonumber \\
&& \prod_{2\le j <k\le N}
\delta\left({1\over 2}[\det_{1,\la j+k \ra}-\det_{1,j} -\det_{1,k}]
-\sqrt{\det_{1,j}\det_{1,k}} I_{j,k}(M-2)\right)
  \label{dist.s3}
\eea
where lemma 2 and $\delta(x/a) = a\;\delta(x)$ are used.

A moment of reflection tells us that $I_{j,k}(M-2)$ is nothing but setting
all the $\theta_{M-1;\, j(k)} = \pi/2$ so that the $M$th component of
vectors $\x_j$ and $\x_k$ are identically zero. Or equivalently, we are
 then looking at vectors living in $M-1$ dimensional space instead of
 $M$ dimensional space. Since the solid angle of vector $\x_1$
 and all the radial components of the vectors are completely
 integrated out, we are now left with $N-1$ {\em unit vectors}  living
 in $M-1$ dimensions.  We can then again require that the $M-1$th
 component of vector $\x_2$ to  be along $\e_{M-1}$ so that $x_2^{M-1} = 1$
 and $x_2^{1\le \alpha\le M-2} = 0$. We then again write the components of
 the other unit vectors in polar angles such as in (\ref{coord.in.polar})
 but with $\theta_{M-1;\, 2\le i\le N} = \pi/2$ and $\theta_{M-2;\, 2}=0$.
 Note that spherical symmetry guarantees that $d\Omega_i(M-1)$ has exactly
 the same form regardless how one chooses the axes.
 Under such consideration, we note again that
 $I_{2,3\le k\le N} = \cos \theta_{M-2;\, k}$ and the delta functions
 again give us that
 \bea
\sqrt{\det_{1,2}\det_{1,k}} \; \cos \theta_{M-2;\, k} &=&
 {1\over 2}[\det_{1,\la 2+k \ra}- \det_{1,2} - \det_{1,k}] \nonumber \\
\sin \theta_{M-2;\, k} &=& \sqrt{1-\cos^2\theta_{M-2;\, k}} =
 \sqrt{\det_{1}\det_{1,2,k} \over \det_{1,2}\det_{1,k}},
\eea
 again we have used $ 0 \le \theta_{M-2;\, k} < \pi$.
 After such understanding, we may then proceed to integrate the polar angles
 $\theta_{M-2;\, 3\le i\le N}$ and, noting that the solid angle integration
 of $d\Omega_2(M-1)$ leads to $K_{M-1}$,  obtain
\bea
P(\{Y_{ij}\})
&=& 2^{-N} \left[\prod_{i=1}^N f_i(Y_{ii}) \right]
 K_M\; K_{M-1}\; \det_{1,2}^{(N-3)(N-M+1)\over 2}
\left[\prod_{j=3}^N \det_{1,2,j}^{M-4\over 2}\right]
\int \left[\prod_{i=3}^{N} d\Omega_i(M-2)\right]
\nonumber \\
&&   \prod_{3\le j <k\le N}
\delta\left({1\over 2}[\det_{1,2,\la j+k \ra}-\det_{1,2,j} -\det_{1,2,k}]
-\sqrt{\det_{1,2,j}\det_{1,2,k}} I_{j,k}(M-3)\right) \nonumber \\
&=& 2^{-N}\left[\prod_{i=1}^N f_i(Y_{ii}) \right]
 K_M\; K_{M-1}\cdots K_{M-L+2}\;
\det_{1,2,\cdots,L-1}^{(N-L)(N-M+1)\over 2}
\left[\prod_{j=L}^N \det_{1,2,\cdots,L-1,j}^{M-L-1\over 2}\right]\nonumber \\
&& \int \left[\prod_{i=L}^{N} d\Omega_i(M-L+1)\right]
\prod_{L\le j <k\le N}
\delta\left({1\over 2}[\det_{1,2,\cdots,L-1,\la j+k \ra}
-\det_{1,2,\cdots,L-1,j} -\det_{1,2,\cdots,L-1,k}] - \right. \nonumber \\
&& \left. \hspace*{0.9in}
-\sqrt{\det_{1,2,\cdots,L-1,j}\det_{1,2,\cdots,L-1,k}}\; I_{j,k}(M-L)\right)
\label{dist.s4}
\eea

Now we see how this process can continue with application of lemmas 1 and 2.
When $N\le M$, the process actually terminate at $L=N$ where all the delta
 functions have been integrated out.
In this way, we have extended the celebrated result of Wishart to the
 more generic case
\bea
P(\{Y_{ij}\})
&=& 2^{-N} \left[\prod_{i=1}^N f_i(Y_{ii}) \right]
\left[\prod_{i=1}^N K_{M-i+1}\right] \;
\det_{1,2,\cdots,N}^{M-N-1\over 2}
 \nonumber\\
&=& \left[\prod_{i=1}^N {f_i(Y_{ii})K_{M-i+1}\over 2}\right] \;
[{\rm det}({\b Y})]^{(M-N-1)/2}
\eea
 and in the more restricted case with $\prod_{i=1}^N f_i(Y_{ii})
 = (2\pi)^{-MN/2}\;\exp({-{1\over 2}{\rm Tr}{\b Y}})$, we have
 exactly the Wishart result
\be
P(\{Y_{ij}\})
= (2\pi)^{-MN/2}\left[\prod_{i=1}^N {K_{M-i+1}\over 2}\right] \;
[{\rm det}({\b Y})]^{(M-N-1)/2}\;\exp({-{1\over 2}{\rm Tr}{\b Y}}).
\ee

For the case of our interest $N>M$, however, the integral does not terminate
that way and there will be leftover delta functions. The furthest we can go
 then is to integrate till $L=M$ together with one last complication that the
 range of angle $\theta_{1;\, i}$ is between $0$ and $2\pi$ (instead of between
 $0$ and $\pi$) and therefore $\sin \theta_{1;\, i}$
can take both positive and  negative signs. To see it explicitly,
we may integrate up to $L=M-1$
 and notice that $I_{j,k}(1) = \cos \theta_{1;\, j} \cos \theta_{1;\, k}
 + \sin \theta_{1;\, j} \sin \theta_{1;\, k}$  because of $I_{j,k}(0) = 1$.
 This way, we have
\bea
P(\{Y_{ij}\})
&=& 2^{-N} \left[\prod_{i=1}^N f_i(Y_{ii}) \right]
 K_M\; K_{M-1}\cdots K_{3}\;
\det_{1,2,\cdots,M-2}^{(N-M+1)(N-M+1)\over 2}
\int \left[\prod_{i=M-1}^{N} d\Omega_i(2)\right]\nonumber \\
&&  \prod_{M-1\le j <k\le N}
\delta\left({1\over 2}[\det_{1,2,\cdots,M-2,\la j+k \ra}
-\det_{1,2,\cdots,M-2,j} -\det_{1,2,\cdots,M-2,k}] - \right. \nonumber \\
&& \left. \hspace*{0.9in}
-\sqrt{\det_{1,2,\cdots,M-2,j}\det_{1,2,\cdots,M-2,k}}\; I_{j,k}(1)\right)
\label{dist.s5}
\eea
We then again choose the effectively the direction of the $(M-1)$th unit
 vector to be along the $\e_{2}$ direction and therefore the solid angle
 of the new unit vector $\x_{M-1}$ that lives in two dimensions.
 This way, we have
\bea
P(\{Y_{ij}\})
&=& 2^{-N}  \left[\prod_{i=1}^N f_i(Y_{ii}) \right]\;
 \left[\prod_{j=2}^{M}K_{j}\right]
\; \det_{1,2,\cdots,M-1}^{(N-M)(N-M+1)\over 2}
\left[\prod_{j=M}^N (\det_{1,2,\cdots,M-1,j})^{-1/2}\right]\nonumber \\
&& \sum_{\rm sgns} \prod_{M\le j <k\le N}
\delta\left({1\over 2}
[\det_{1,2,\cdots,M-1,\la j+k \ra}
-\det_{1,2,\cdots,M-1,j} -\det_{1,2,\cdots,M-1,k}] - \right.
\nonumber \\
&& \left.  \hspace*{0.3in}
- {\rm sgns}\;
\;\sqrt{\det_{1,2,\cdots,M-1,j} \; \det_{1,2,\cdots,M-1,k}} \; \right)
\label{dist.s6}
\eea
where
\be
\int_0^{2\pi} d\theta = \int_0^\pi -{d\cos \theta \over \sin \theta}
+ \int_\pi^{2\pi} -{d\cos \theta \over \sin \theta} =
{1\over |\sin\theta|}
\left[\int_{-1}^{1} d\cos \theta \;\large|_{\sin\theta \ge 0}
 + \int_{-1}^{1} d\cos \theta \;\large|_{\sin\theta \le 0}
\right]
\ee
is used.
Now the sum over ``sgns'' deserves some explanations.  Each candidate
of $\sin\theta_{1;\, i}$ in eq.(\ref{dist.s5})
 in principle can take both positive and negative
values. This means that for each $k$, the quantity $\sqrt{\det_{1,2,\cdots,
M-1, k}}$ can carry both positive and negative signs. Because each of
the remaining  reduced unit vectors $\x_M, \x_{M+1}, \cdots,\x_N$  can
play a role in $\sqrt{\det_{1,2,\cdots,M-1, k}}$, we therefore have to
consider $2^{N-M+1}$ different combinations. That is to say, our
sum over signs actually consists of $2^{N-M+1}$ terms each of which
 is a product of $(N-M)(N-M+1)/2$ delta functions. In order to better organize
 these delta functions, we introduce two new notations:
\bea
b_k &\equiv & \sqrt{\det_{1,2,\cdots,M-1,k}}  \\
B_{k,l}&\equiv& {1 \over 2}
[b^2_{\la k+l \ra} -b^2_k -b^2_l ].
\eea
Under this new notation, we may rewrite our distribution function as
\bea
P(\{Y_{ij}\})
&=& 2^{-N} \left[\prod_{i=1}^N f_i(Y_{ii}) \right]
 \; \left[\prod_{j=2}^{M}K_{j}\right]
\; \det_{1,2,\cdots,M-1}^{(N-M)(N-M+1)\over 2}
\left[\prod_{j=M}^N (\det_{1,2,\cdots,M-1,j})^{-1/2}\right]\nonumber \\
&& \sum_{ \{ s_i = \pm 1\} } \prod_{M\le j <k\le N}
\delta\left(B_{jk}- s_j s_k \;b_j\; b_k \right),
\label{dist.s7}
\eea
 where $s_i = \pm 1$ are Ising variables conveniently introduced to
 represent the signs needed. Note that although
 eq.(\ref{dist.s7}) could be regarded as the end result of integrations,
 to render it useful we will reassemble these $2^{N-M+1}$ combinations
 into a single term. This is in some way similar to obtain the partition
 function of an Ising system by summing up all possible spin
 configurations. We may also say that
 obtaining eq.(\ref{dist.s7}) is only half way to our  goal.

 To work towards the final goal, we now start the task of reassembling
these $2^{N-M+1}$ terms of product of delta functions.
When applied to symmetric matrices, lemma 1 (with $L \to M$) tells us that
\be
b^2_k b^2_l - B^2_{k,l} = \det_{1,2,\cdots,M-1}\;\cdot\;
\det_{1,2,\cdots,M-1,k,l}.
\ee
 With the labeling of $k\in \{M, M+1, \cdots, N\}$,
 we can now order the $\pm$ signs carried by each $b_k$ in the following
manner. First, we observe that if we change the signs of every $b_k$,
 the delta function is invariant. This immediately leads to a two fold symmetry
 which allows us to require that the sign carried by $b_N$ being always
 positive. Although there are in total $2^{N-M+1}$ terms in the sum, the
 two-fold symmetry dictates only $2^{N-M}$ different terms. These
 terms are selected by our choosing $b_N$ always carrying positive sign, i.e.
 $s_N = 1$.
 For the $2^{N-M}$ different terms, we generate them in the follows:
 We first write down the two cases where $b_M$ can be either positive or
 negative. We organize it as
\be
\begin{array}{c|c}
 & s_M\\
\noalign{\hrule}
& + \\
& - \end{array}
\ee
We then make two identical such copies. For the first copy
 we have $b_{M+1}$ carrying positive signs, and for the second copy
 we have $b_{M+1}$ carrying negative signs.
\be
\begin{array}{c|cc}
  & s_{M+1} & s_M\\
\noalign{\hrule}
& + & +\\
& + & - \\
& - & + \\
& - & - \end{array}
\ee
We then make two identical copies of the above. We again in the first copy
 put in $b_{M+2}$ that carries positive signs and in the second copy
 put in $b_{M+2}$ that carries negative signs. We then arrives at
\be
\begin{array}{c|ccc}
  & s_{M+2} & s_{M+1} & s_M \\
\noalign{\hrule}
& + & + & +\\
& + & + & - \\
& + & - & + \\
& + & - & - \\
& - & + & +\\
& - & + & - \\
& - & - & + \\
& - & - & -
\end{array}
\ee
 This process keeps going till we adding positive-sign-carrying $b_{N-1}$
 to first copy and negative-sign-carrying $b_{N-1}$ to the second copy.
 After that we add positive-sign-carrying $b_N$ to every term. We finally
 have something look like
\be
\begin{array}{c|cccccc}
  & s_N & s_{N-1} &\cdots& s_{M+2}  & s_{M+1} & s_M \\
\noalign{\hrule}
1 & + & + & \cdots & + & + & +\\
2 & + & + & \cdots & + & + & - \\
3 & + & + & \cdots & + & - & + \\
4 & + & + & \cdots & + & - & - \\
5 & + & + & \cdots & - & + & +\\
6 & + & + & \cdots & - & + & - \\
7 & + & + & \cdots & - & - & + \\
8 & + & + & \cdots & - & - & - \\
\vdots  & \vdots & \vdots & \ddots & \vdots & \vdots & \vdots \\
2^{N-M}-3 & + & - & \cdots & - & + & + \\
2^{N-M}-2 & + & - & \cdots & - & + & - \\
2^{N-M}-1 & + & - & \cdots & - & - & + \\
2^{N-M}   & + & - & \cdots & - & - & -
\end{array}\label{arr}
\ee
We now start by combining terms $2l-1$ and $2l$ for all $l \le 2^{N-M-1}$.
In order to simplify the notation, we shall only write out the
$\sum_{\{ s_i = \pm 1 \}}
\prod_{M\le j<k\le N}\delta(B_{jk}- s_j s_k\;b_j\;b_k)$ part and
multiply the final results by appropriate factor later.
In table (\ref{arr}), the first term has all the $b_k$ carrying
 positive signs, i.e. $s_k = 1$,
while the second term has $b_M$ carrying negative sign
 but with the rest of $b_k$ carrying positive signs. The sum of the
first and the second term therefore look like
\bea
&& \left[\delta(B_{M,M+1}-b_M\;b_{M+1})
\prod_{k=M+2}^N\delta(B_{M,k} -b_M\;b_k) \right. \nonumber \\
&& \;\;\; \left. + \;\delta(B_{M,M+1} +
 b_M\;b_{M+1}) \prod_{k=M+2}^N\delta(B_{M,k} +b_M\;b_k)
 \right]\prod_{(M+1)\le k<l\le N} \delta(B_{k,l}-b_k\;b_l)
\eea
In the $\prod_{k=M+2}^N \delta(B_{M,k} -b_M\;b_k)$ part of the
 first term, we replace $b_M$ by $B_{M,M+1}/b_{M+1}$ and $b_k$ by
 $B_{M+1,k}/b_{M+1}$. For the  $\prod_{k=M+2}^N \delta(B_{M,k}+b_M\;b_k)$ part
of second term, we replace $b_M$ by $-B_{M,M+1}/b_{M+1}$ and $b_k$ by
 $B_{M+1,k}/b_{M+1}$. This way, the sum of the term 1 and term 2 can be
 rewritten as
 \bea
&& \left[\delta(B_{M,M+1}-b_M\;b_{M+1})
 + \;\delta(B_{M,M+1} + b_M\;b_{M+1}) \right] \nonumber \\
&&\;\;\;\;\
  \prod_{k=M+2}^N\delta(B_{M,k} - {B_{M,M+1}B_{M+1,k}\over b^2_{M+1}} )
 \prod_{(M+1)\le k<l\le N} \delta(B_{k,l}-b_k\;b_l)\nonumber \\
&& \;\; = 2 { b_{M}\;b_{M+1} \over \det_{1,2,\cdots, M-1}}\;
\delta(\det_{1,2\cdots,M-1,M,M+1})  \nonumber \\
&& \;\;\;\;\
  \prod_{k=M+2}^N\delta(B_{M,k} - {B_{M,M+1}B_{M+1,k}\over b^2_{M+1}} )
 \prod_{(M+1)\le k<l\le N} \delta(B_{k,l}-b_k\;b_l)
\eea
Similarly, the third and the fourth terms can also be added to
\bea
&& \left[\delta(B_{M,M+1}+b_M\;b_{M+1})
\prod_{k=M+2}^N\delta(B_{M,k} -b_M\;b_k)\;\delta(B_{M+1,k} -b_{M+1}\;b_k)
 \right. \nonumber \\
&& \;\;\; \left. + \;\delta(B_{M,M+1} -
 b_M\;b_{M+1}) \prod_{k=M+2}^N\delta(B_{M,k} + b_M\;b_k)\;
\delta(B_{M+1,k} +b_{M+1}\;b_k)
 \right] \nonumber \\
&& \;\;\; \prod_{(M+1)\le k<l\le N} \delta(B_{k,l}-b_k\;b_l),
\eea
and with similar reasoning this sum 
 leads to
 \bea
&& 2 { b_{M}\;b_{M+1} \over \det_{1,2,\cdots, M-1}}\;
\delta(\det_{1,2\cdots,M-1,M,M+1}) \;
  \prod_{k=M+2}^N\delta(B_{M,k} - {B_{M,M+1}B_{M+1,k}\over b^2_{M+1}} )
 \nonumber \\
&& \;\;\;\;\
  \prod_{k=M+2}^N\delta(B_{M+1,k} + b_{M+1}\; b_k )
 \prod_{(M+2)\le k<l\le N} \delta(B_{k,l}-b_k\;b_l).
\eea
One thing we notice immediately is that if we factor out the
common factor between the sum of term 1 and 2 as well as the sum of term 3
 and 4, we see an expression that is very similar to the original one as if
 the variable $b_M$ does not come into the picture in the first place.

In fact the common factor
\be
2 { b_{M}\;b_{M+1} \over \det_{1,2,\cdots, M-1}}\;
\delta(\det_{1,2\cdots,M-1,M,M+1}) \;
  \prod_{k=M+2}^N\delta(B_{M,k} - {B_{M,M+1}B_{M+1,k}\over b^2_{M+1}} )
\ee
is the same for every pairwise sum of $k=2l-1$ and $k=2l$.
The reason is very simple. Originally we have for $M+2 \le k \le N$
\be \label{subst}
\delta \left(B_{M,k}-s_M s_k \;b_M\; b_k \right)
\ee
in the product and we also have delta functions
\bea
\delta \left(B_{M,M+1}- s_M s_{M+1}\; b_M\;b_{M+1}
\right) \nonumber\\
\delta \left(B_{M+1,k}- s_{M+1}s_k\; b_{M+1}\; b_k \right)
\eea
which give us
\be
s_M b_M = {B_{M,M+1}\over s_{M+1}\; b_{M+1}} \;\;\;\;
{\rm and} \;\;\;\;
s_k b_k = {B_{M+1,k}\over s_k \; b_{k}}
\ee
Upon substitution into eq.(\ref{subst}) we have
\be \label{subst.1}
\delta \left(B_{M,k}-{ B_{M,M+1} B_{M+1,k}\over
(s_{M+1})^2b^2_{M+1}} \right)
\ee
and using the fact that $(s_k)^2 = 1$ for all $1\le k\le N$. We
thus proved that every pairwise sum produce such common factor.
 Furthermore, when we sum up those terms together, we reduced the number of
 terms exactly by half and at the same time gain a factor $2$.

Remember that the sign of $b_N$ is fixed to be positive, i.e.
 $s_N = +1 $ and this process ends when turning
$\delta (B_{N-1,N}- s_{N-1} b_{N-1} b_N) +
\delta (B_{N-1,N}+ s_{N-1} b_{N-1} b_N ) $ into a single
delta function, which corresponds to
$k=N-1$ and $l=N$. Note that we start with $k=M$, therefore, we
have to do the same process for $N-M$ times.
Together with the two-fold symmetry we mentioned at the very beginning,
we will end up with
\be
2\cdot 2^{N-M}
\left[\prod_{j=M}^{N-1} {b_j\; b_{j+1} \over \det_{1,2,\cdots,M-1}}
\delta(\det_{1,2,\cdots,M-1,j,j+1})\right] \prod_{j=M}^{N-2}\prod_{k=j+2}^N
\delta\left( B_{j,k}- {B_{j,j+1} B_{j+1,k}\over b_{j+1}^2} \right)
\ee
To further simplify the expression, we see from lemma 2 that
\be
B_{j,l}B_{l,k}-b_l^2B_{j,k} = -{1\over 2}(\det_{1,2,\cdots, M-1,l,\la j+k\ra}
 - \det_{1,2,\cdots,M-1,l,j}-\det_{1,2,\cdots,M-1,l,k})\cdot
\det_{1,2,\cdots, M-1}
\ee
and with $l\to j+1$, we see that
\bea
&&\delta(B_{j,k}-{B_{j,j+1}\;B_{j+1,k}\over b^2_{j+1}}) \nonumber \\
&&\;\;\;\; = {\det_{1,2,\cdots,M-1,j+1}\over \det_{1,2,\cdots,M-1}}
\delta\left({1\over 2}[\det_{1,2,\cdots,M-1,j+1,\la j+k\ra}
-\det_{1,2,\cdots,M-1,j+1,j} -\det_{1,2,\cdots,M-1,j+1,k}]\right)
\eea
We can now include the part we did not include explicitly and obtain finally
\bea
P(\{Y_{ij}\})
&=& \left[\prod_{i=1}^N f_i(Y_ii) \right]\;
\left[\prod_{j=1}^{M}{K_{j}\over 2}\right] \prod_{j=M+1}^{N-1}
\det_{1,2,\cdots,M-1,j}^{N-j-{1\over 2}}\;\;
\prod_{j=M}^{N-1} \delta(\det_{1,2,\cdots,M-1,j,j+1})\nonumber \\
&& \!\! \prod_{j=M}^{N-2} \prod_{k=j+2}^N
\delta\left({1\over 2}[\det_{1,2,\cdots,M-1,j+1,\la j+k\ra}
-\det_{1,2,\cdots,M-1,j+1,j} -\det_{1,2,\cdots,M-1,j+1,k}]\right).
\eea

The reason that this final expression does not look very symmetric
 come from the fact that when we exploited the gauge degrees of freedom
 we chose  $\x_1$ to be parallel to $\hat e_M$, $\x_2$ parallel to
 $\hat e_{M-1}$ etc. Therefore, any permutation of the vectors
 in the order of integration should give identical results.
 When one consider that version, the symmetry will appear explicitly.
 The current expression, however, could be more useful from
the standpoint of numerical use as will be explained in a separate
 publication~\cite{YZ2} where more details will be presented.
An alternative way to get the equivalent expression is to
 diagonalize the $\b Y$ matrix first, and then focus on
 the nonzero eigenvalues of  the $\b Y$ matrix.
Being  seemingly  more elegant, this way gives the same results and does
 not necessarily provide an easier way to calculate conditional probabilities
 for predicting redundant information. The application of our results
as well as comparison of various approaches will be discussed
in a later publication~\cite{YZ2}.
Because our method only assumes spherical symmetry in the
ensemble of vectors, it can be applied to magnetic systems where
 the spins are of fixed lengths. Our method also bears potential to
 model the real knowledge network where the components of a vector might
 not be completely independent random Gaussian variables.

 Finally, let us end with a brief note. After the calculational
 part of this work was completed,
 Janik and Nowak~\cite{polish} recently presented a similar result
 which followed closely the route in~\cite{Wilks}.
 Their method is simple but crucially depends on the random Gaussian ensemble
 assumed. Interestingly, the two results display a crucial
 discrepency in terms of the number of constraints (delta functions).
 Although the degrees of freedom counting is
 somehow elementary, we would like to go over it again here.
 We first note that there are $NM$ integration variables.
 However, the gauge symmetry assures that only $NM- [M(M-1)/2]$ of those
 integrations are independent with respect to the $\b Y$ matrix.
 Since the $\b Y$ matrix has $N(N+1)/2$ independent matrix elements,
 we therefore find that the final number of delta functions should be
 $N(N+1)/2- \{NM- [M(M-1)/2]\} = (N-M)(N-M+1)/2$. This expected number
 of delta functions indeed show up naturally in our final expression.

\section*{Acknowledgement}
We thank the hospitality of ITP at UCSB where this work was initiated.
 YKY wishes to thank V.P. Nair and R. Friedberg for useful discussions and
encouragement. He also acknowledge the hospitality of Center for Studies
 in Physics and Biology of Rockefeller University where the calculational
 part of this  work was  completed.
This work is supported in part by NSF DMR-0110903.

\centerline{APPENDIX}

\noindent Let us first prove lemma 1. \\
For illustration purpose, we
draw the matrix determinant represented by
$\det_{1,2,\cdots, L-1,\la k+j\ra}$ as
\be
{\rm det} \left({\begin{array}{c|c|c|c|c}
        Y_{11}&Y_{12} &\cdots & Y_{1,L-1} &Y_{1k}+Y_{1j}\\
        \noalign{\hrule}
        Y_{21}&Y_{22} &\cdots & Y_{2,L-1} &Y_{2k}+Y_{2j}\\
        \noalign{\hrule}
        \vdots & \vdots &\ddots & \vdots & \vdots\\
        \noalign{\hrule}
        Y_{L-1,1}& Y_{L-1,2}& \cdots & Y_{L-1,L-1} &Y_{L-1,k}+Y_{L-1,j} \\
        \noalign{\hrule}
        Y_{k1}+Y_{j1} & Y_{k2}+Y_{j2}&\cdots&Y_{k,L-1}+Y_{j,L-1}&
        Y_{kk}+Y_{jj}+Y_{kj}+Y_{jk}
	\end{array}} \right). \nonumber
\ee

From above illustration, we have
\bea
\det_{1,2,\cdots, L-1,\la k+j\ra} &=& [Y_{kk}+Y_{jj}+Y_{jk}+Y_{kj}]
\;\det_{1,2,\cdots,L-1} \nonumber\\
&&\hspace*{-0.8in} + \sum_{n=1}^{L-1}(-1)^{L+n}  [Y_{kn}
+Y_{jn}]\tA_{L,n}(\la k+j\ra )\label{det.decomp.k+j}
\eea
where $\tA_{L,n}(\la k+j\ra )$
denotes the minor of the matrix element on row $L$ column $n$
in the matrix shown above. Note that the elements on the $L$th column
 are given by $Y_{n \in \{1,2,\cdots, L-1\}, \la k+j \ra }$ and the
 elements in the $L$th row are given by
$Y_{\la k+j \ra ,n \in \{1,2,\cdots, L-1\}}$.
Similarly,
\bea
\det_{1,2,\cdots, L-1,k} &=& \sum_{n=1}^{L-1}Y_{kn}
(-1)^{L+n}\tA_{L,n}(k) \nonumber \\
&& + Y_{kk}\; \det_{1,2,\cdots,L-1} \label{det.decomp.k}
\eea
and an identical expression exist for $\det_{1,2,\cdots, L-1,j}$.
One immediate observation is that
\be
\tA_{L,n}(\la k+j \ra) =
 \tA_{L,n}(k) + \tA_{L,n}(j).
\ee
 This part can be easily seen if we calculate each minor by
expanding along its  last column $L-1$: summing  elements on row $L-1$
and column $1\le n\le L-1$ multiply by their corresponding subminors.
 Note that
\be
\tA_{L,n}(\la k+j\ra) = {\rm det} \;
\left({\begin{array}{ccccccc}
        Y_{11}&Y_{12} &\cdots & Y_{1,n-1} & Y_{1,n+1}&\cdots&Y_{1k}+Y_{1j}\\
        Y_{21}&Y_{22} &\cdots & Y_{2,n-1} & Y_{2,n+1}&\cdots&Y_{2k}+Y_{2j}\\
        \vdots & \vdots & & \vdots & \vdots & & \vdots\\
        Y_{L-1,1}& Y_{L-1,2}& \cdots & Y_{L-1,n-1} &Y_{L-1,n+1}&
          \cdots &Y_{L-1,k}+Y_{L-1,j}
	\end{array}} \right),
\ee
and
\be
\tA_{L,n}(k) = {\rm det} \;
\left({\begin{array}{ccccccc}
        Y_{11}&Y_{12} &\cdots & Y_{1,n-1} & Y_{1,n+1}&\cdots&Y_{1k}\\
        Y_{21}&Y_{22} &\cdots & Y_{2,n-1} & Y_{2,n+1}&\cdots&Y_{2k}\\
        \vdots & \vdots & & \vdots & \vdots & & \vdots\\
        Y_{L-1,1}& Y_{L-1,2}& \cdots & Y_{L-1,n-1} &Y_{L-1,n+1}&
          \cdots &Y_{L-1,k}
	\end{array}} \right).
\ee
Since the subminors associated with
such decomposition are exactly the same whether the elements at
the last column is $\la k+j\ra$ or just $k$ or $l$.

Now let us re-express the quantity of interest
\bea
&& \det_{1,2,\cdots,L-1,\la k+j\ra} -
\det_{1,2,\cdots, L-1,k}-\det_{1,2,\cdots, L-1,j} \nonumber \\
&& \;\; = \sum_{n=1}^{L-1} (-1)^{L+n}\left[Y_{kn}(\tA_{L,n}(\la k + j \ra)
   \right. \nonumber \\
&& \hspace*{0.7in}\left.
 - \tA_{L,n}(k) + Y_{jn}(\tA_{L,n}(\la k+j\ra)-\tA_{L,n}(j)\right] \nonumber \\
&& \;\;\;\hspace*{0.3in} + [Y_{kj}+Y_{jk}]
 \cdot \det_{1,2,\cdots,L-1} \nonumber \\
&& \;\; = \sum_{n=1}^{L-1} (-1)^{L+n}\left[Y_{kn}\tA_{L,n}(j)
 + Y_{jn}\tA_{L,n}(k)\right] \nonumber \\
&& \;\;\;\hspace*{0.3in} + [Y_{kj}+Y_{jk}] \cdot \det_{1,2,\cdots,L-1}
\label{det.k+j-det.k-det.j}
\eea
And consequently, we have the square of the above  expression as
\bea
&&[\det_{1,2,\cdots,L-1,\la k+j\ra} -
\det_{1,2,\cdots, L-1,k}-\det_{1,2,\cdots, L-1,j}]^2 \nonumber \\
&& \;\;\; = \;[Y_{jk}+Y_{kj}]^2\cdot \det^2_{1,2,\cdots, L-1}
 + 2\;[Y_{jk}+Y_{kj}]\cdot \det_{1,2,\cdots,L-1}
\sum_{n=1}^{L-1} (-1)^{L+n}\left[Y_{kn}\tA_{L,n}(j) + Y_{jn}\tA_{L,n}(k)
\right] \nonumber \\
&& \;\;\;\;\
+ \{\sum_{n=1}^{L-1} (-1)^{L+n}\left[Y_{kn}\tA_{L,n}(j)
+ Y_{jn}\tA_{L,n}(k)\right] \}^2 \label{det.L.k+j-k-j.square}
\eea

Using eq.(\ref{det.decomp.k}) and a similar one with $k$ replaced by $j$, we
 now write
\bea
&&\det_{1,2,\cdots,L-1,k} \cdot \det_{1,2,\cdots,L-1,j} \nonumber \\
&& \;\;\;\;\;\; = \sum_{n=1}^{L-1}\sum_{n'=1}^{L-1} Y_{kn}\tA_{L,n}(k)
 Y_{jn'}\tA_{L,n'}(j) \nonumber\\
&&\;\;\;\;\;\;\;\;\; + Y_{kk}\cdot \det_{1,2,\cdots,L-1}
\sum_{n'=1}^{L-1}(-1)^{L+n'}Y_{jn'}\tA_{L,n'}(j)
+Y_{jj}\cdot \det_{1,2,\cdots,L-1}
\sum_{n=1}^{L-1}(-1)^{L+n}Y_{kn}\tA_{L,n}(k) \nonumber \\
&& \;\;\;\;\;\;\;\;\; + Y_{jj}Y_{kk}\cdot \det^2_{1,2,\cdots,L-1}
\eea

We are now just one step away from proving lemma 1. We start with
a determinant identity
\be
{\bf Y} = \left({\begin{array}{ccc|c}
        &   &  &    \\
        & A &  & C  \\
        &   &  &    \\
        \noalign{\hrule}
        & D &  & B
	\end{array}} \right) =
\left({\begin{array}{ccc|c}
        &   &  &    \\
        & A &  & 0  \\
        &   &  &    \\
        \noalign{\hrule}
        & D &  & I_M
	\end{array}} \right)
\left({\begin{array}{ccc|c}
        &   &  &    \\
        & I_N &  & A^{-1}C  \\
        &   &  &    \\
        \noalign{\hrule}
        & 0 &  & B-DA^{-1}C
	\end{array}} \right)
\ee
and consequently
\be
{\rm det}({\bf Y}) = {\rm det}(A) \cdot {\rm det}(B-DA^{-1}C).
\ee
This tells us that we can rewrite $\det_{1,2,\cdots,L-1,k,j}$ in the
following form:
\be
\det_{1,2,\cdots,L-1,k,j}\;=\; {\rm det}
\left({\begin{array}{ccc|cc}
       & & & Y_{1k} & Y_{1j} \\
       & \left({\b Y}\right)_{L-1} & &\vdots &\vdots \\
       &  & & Y_{L-1\;k} &Y_{L-1\;j}    \\
	\noalign{\hrule}
     Y_{k1}& Y_{k2} & \cdots & Y_{kk} & Y_{kj}  \\
     Y_{j1}& Y_{j2} & \cdots & Y_{jk} &Y_{jj}
	\end{array}} \right)
= \det_{1,2,\cdots,L-1}\; \cdot\; {\rm det}(B-D{({\b Y})_{L-1}}^{-1}C)
\label{det.k.j}
\ee
where
\be
B = \left(\begin{array}{cc}
       Y_{kk} & Y_{kj}  \\
	Y_{jk} &Y_{jj}
	\end{array}\right), \hspace*{0.2in}
D = \left(\begin{array}{cccc}
	Y_{k1} & Y_{k2} & \cdots & Y_{k\; L-1}\\
	Y_{j1} & Y_{j2} & \cdots & Y_{j\; L-1}
        \end{array} \right)
\hspace*{0.2in} {\rm and} \hspace*{0.2in}
C  = \left(\begin{array}{cc}
        Y_{1k} & Y_{1j} \\
	Y_{2k} & Y_{2j} \\
        \vdots & \vdots \\
        Y_{L-1\; k} & Y_{L-1\; j}
	\end{array}\right)
\ee

We see that the matrix  $G \equiv D({\b Y}^{-1})_{L-1}C$ have its components as
\bea
G_{11} &=& \sum_{n=1}^{L-1}\sum_{n'=1}^{L-1} Y_{kn}\Gamma_{n,n'}Y_{n'k}
\nonumber \\
G_{12} &=& \sum_{n=1}^{L-1}\sum_{n'=1}^{L-1} Y_{kn}\Gamma_{n,n'}Y_{n'j}
\nonumber \\
G_{21} &=& \sum_{n=1}^{L-1}\sum_{n'=1}^{L-1} Y_{jn}\Gamma_{n,n'}Y_{n'k}
\nonumber \\
G_{22} &=& \sum_{n=1}^{L-1}\sum_{n'=1}^{L-1} Y_{jn}\Gamma_{n,n'}Y_{n'j}
\eea

Note that $\Gamma = {\b Y}^{-1}$ is the inverse of the matrix
\be
{\b Y} = \left(\begin{array}{ccc}
    Y_{11} & \cdots & Y_{1\; L-1} \\
    \vdots &        & \vdots \\
    Y_{L-1\;1} & \cdots & Y_{L-1\;L-1}
    \end{array} \right)
\ee
One thing to remember is that
\be
\Gamma_{n,n'} = (-1)^{n+n'}H_{n',n}/\det_{1,2,\cdots,L-1}
\ee
where $H_{n',n}$ is the minor corresponding to matrix element
 at row $n'$ and column $n$ of the matrix $({\b Y})_{L-1}$.

The most important observation here is that
\be
\sum_{n'=1}^{L-1}\Gamma_{n,n'}Y_{n'j} =
\sum_{n'=1}^{L-1}(-1)^{n+n'}Y_{n'j}H_{n',n}/\det_{1,2,\cdots,L-1}
\ee
Note that the RHS of the above eq is nothing but expanding the determinant
of the following matrix along the $n$th column
\be
\left(\begin{array}{ccccccc}
Y_{11}&\cdots&Y_{1\;n-1}&Y_{1j} & Y_{1\;n+1} & \cdots &Y_{1\;L-1}\\
Y_{21}&\cdots&Y_{2\;n-1}&Y_{2j} & Y_{2\;n+1} & \cdots &Y_{2\;L-1}\\
\vdots &     &\vdots    &\vdots & \vdots     &        &  \vdots \\
Y_{L-1\;1} & \cdots &Y_{L-1\;n-1}&Y_{L-1\;j} &Y_{L-1\;n+1} &\cdots &
 Y_{L-1\; L-1} \end{array}\right)
\ee
divided by $\det_{1,2,\cdots,L-1}$. Moreover,
\be
{\rm det}\left(\begin{array}{ccccccc}
Y_{11}&\cdots&Y_{1\;n-1}&Y_{1j} & Y_{1\;n+1} & \cdots &Y_{1\;L-1}\\
Y_{21}&\cdots&Y_{2\;n-1}&Y_{2j} & Y_{2\;n+1} & \cdots &Y_{2\;L-1}\\
\vdots &     &\vdots    &\vdots & \vdots     &        &  \vdots \\
Y_{L-1\;1} & \cdots &Y_{L-1\;n-1}&Y_{L-1\;j} &Y_{L-1\;n+1} &\cdots &
 Y_{L-1\; L-1} \end{array}\right) = (-1)^{L+n-1}{\tA}_{L,n}(j)
\ee
by the definition of ${\tA}_{L,n}(j)$. This is due to the fact that
we need to switch columns $(L-1)-n$ times in order to
move the rightmost column $L-1$ to be at column $n$ without changing the
order of the rest. And also because $(-1)^{2n} =1 $ always.

We therefore have
\bea
\sum_{n'=1}^{L-1} \Gamma_{n,n'}Y_{n'j} &=& (-1)^{L+n-1}\tA_{L,n}(j) /
 \det_{1,2,\cdots,L-1} \nonumber \\
\sum_{n=1}^{L-1} Y_{jn}\Gamma_{n,n'} &=& (-1)^{L+n'-1}\tA_{n',L}(j) /
 \det_{1,2,\cdots,L-1}
\eea
Note that the second equality came from transposing the first equality
 and then set the dummy variables $n' \to n$ and call $n$ $n'$.
Since the determinant is invariant under matrix transposition, the RHS
 is obtained by just swapping $n$ and $n'$ and also transposing the minor.
 Note also that $\tA_{n',L}(j)$ is the minor of
 the matrix element on row $n'$ column $L$ of the matrix ${\b Y}_{L-1}$.

We therefore can rewrite the matrix element $G$ in a much simpler fashion
\bea
G_{11} &= \sum_{n=1}^{L-1} (-1)^{L+n-1} Y_{kn}\tA_{L,n}(k)
 / \det_{1,2,\cdots,L-1} & =
\sum_{n=1}^{L-1} (-1)^{L+n-1} Y_{nk}\tA_{n,L}(k)/ \det_{1,2,\cdots,L-1}
\nonumber \\
G_{12} &= \sum_{n=1}^{L-1} (-1)^{L+n-1} Y_{kn}\tA_{L,n}(j)
 / \det_{1,2,\cdots,L-1} &=
\sum_{n=1}^{L-1} (-1)^{L+n-1} Y_{nj}\tA_{n,L}(k)/ \det_{1,2,\cdots,L-1}
\nonumber \\
G_{21} &= \sum_{n=1}^{L-1} (-1)^{L+n-1} Y_{jn}\tA_{L,n}(k)
 / \det_{1,2,\cdots,L-1} &=
\sum_{n=1}^{L-1} (-1)^{L+n-1} Y_{nk}\tA_{n,L}(j)/ \det_{1,2,\cdots,L-1}
\nonumber \\
G_{22} &= \sum_{n=1}^{L-1} (-1)^{L+n-1} Y_{jn}\tA_{L,n}(j)
 / \det_{1,2,\cdots,L-1} &=
\sum_{n=1}^{L-1} (-1)^{L+n-1} Y_{nj}\tA_{n,L}(j)/ \det_{1,2,\cdots,L-1}
\label{Gmat}
\eea

We therefore may write $\det_{1,2,\cdots,L-1}\cdot \det_{1,2,\cdots,L-1,k,j}$
as
\bea
&& \det_{1,2,\cdots,L-1}\cdot \det_{1,2,\cdots,L-1,k,j}  =
\det_{1,2,\cdots,L-1}^2 \cdot {\rm det}(B-G) \\
&&\;\;\;\; = \left[Y_{kk}\cdot \det_{1,2,\cdots, L-1}
 +\sum_{n=1}^{L-1}(-1)^{L+n}Y_{kn}\tA_{L,n}(k) \right]
\left[Y_{jj}\cdot \det_{1,2,\cdots,L-1} + \sum_{n=1}^{L-1}
 (-1)^{L+n}Y_{jn}\tA_{L,n}(j)\right] \nonumber \\
&&\;\;\;\;\; - \left[Y_{kj}\cdot \det_{1,2,\cdots, L-1}
 +\sum_{n=1}^{L-1}(-1)^{L+n}Y_{kn}\tA_{L,n}(j) \right]
\left[Y_{jk}\cdot \det_{1,2,\cdots,L-1} + \sum_{n=1}^{L-1}
 (-1)^{L+n}Y_{jn}\tA_{L,n}(k)\right]\nonumber
\eea

After some algebra and using eqs(\ref{det.decomp.k})
and (\ref{det.L.k+j-k-j.square}), we can write down the expression explicitly
\bea
&& 4\;\det_{1,2,\cdots,L-1,k}\cdot \det_{1,2,\cdots,L-1,j} -
\left( \det_{1,2,\cdots,L-1,\la k+j\ra}
-\det_{1,2,\cdots,L-1,k}-\det_{1,2,\cdots,L-1,j}\right)^2
\nonumber \\
&& \hspace*{0.6in} - 4\; \det_{1,2,\cdots,L-1} \cdot \det_{1,2,\cdots,
L-1,k,j} \nonumber \\
&& \;\;\; = -(Y_{jk}-Y_{kj})^2\cdot \det_{1,2,\cdots,L-1}^2
-\left[ \sum_{n=1}^{L-1} (-1)^{L+n}\left(Y_{kn}\tA_{L,n}(j)
 -Y_{jn}\tA_{L,n}(k)\right)\right]^2 \nonumber\\
&& \;\;\;\;\;\;\;+ 2Y_{jk}\cdot\det_{1,2,\cdots,L-1}
\left[ \sum_{n=1}^{L-1} (-1)^{L+n}\left(Y_{kn}\tA_{L,n}(j)
 -Y_{jn}\tA_{L,n}(k)\right)\right] \nonumber \\
&& \;\;\;\;\;\;\;- 2Y_{kj}
\left[ \sum_{n=1}^{L-1} (-1)^{L+n}\left(Y_{kn}\tA_{L,n}(j)
 -Y_{jn}\tA_{L,n}(k)\right)\right] \nonumber\\
&& \;\;\; = -\left\{ (Y_{kj}-Y_{jk})\cdot\det_{1,2,\cdots,L-1}
+ \left[ \sum_{n=1}^{L-1} (-1)^{L+n}\left(Y_{kn}\tA_{L,n}(j)
 -Y_{jn}\tA_{L,n}(k)\right)\right]\right\}^2 \nonumber \\
&& \;\;\; = -\left[ {\rm det}\left({\b Y}_{L-1}(j_{L_c},k_{L_r})\right)-
 {\rm det}\left({\b Y}_{L-1}(k_{L_c},j_{L_r})\right) \right]^2
\label{theorem1}
\eea
where the matrices ${\b Y}_{L-1}(j_{L_c},k_{L_r})$ and
${\b Y}_{L-1}(k_{L_c},j_{L_r})$ look like
\be
{\b Y}_{L-1}(j_{L_c},k_{L_r}) =
   \left({\begin{array}{ccc|c}
       & & & Y_{1j} \\
       & \left({\b Y}\right)_{L-1} & &\vdots \\
       &  & & Y_{L-1\;j}    \\
	\noalign{\hrule}
     Y_{k1}& Y_{k2} & \cdots & Y_{kj}
	\end{array}} \right)
 \hspace*{0.2in}
{\b Y}_{L-1}(j_{L_c},k_{L_r}) =
   \left({\begin{array}{ccc|c}
       & & & Y_{1k} \\
       & \left({\b Y}\right)_{L-1} & &\vdots \\
       &  & & Y_{L-1\;k}    \\
	\noalign{\hrule}
     Y_{j1}& Y_{j2} & \cdots & Y_{jk}.
	\end{array}} \right)
\ee

Note that if the matrix elements of $Y$ is symmetric, i.e. $Y_{jk} = Y_{kj}$,
 the RHS of eq.(\ref{theorem1}) is identically zero. This is because the
 inverse of a symmetric matrix will also be symmetric. Therefore
\bea
\sum_{n=1}^{L-1} (-1)^{L+n-1} Y_{kn}\tA_{L,n}(j) &=&
\sum_{n,n'} Y_{kn}\Gamma_{n,n'}Y_{n'j}
= \sum_{n,n'} Y_{nk}\Gamma_{n',n}Y_{jn'} = \sum_{n,n'}Y_{jn'}
\Gamma_{n',n} Y_{nk} \nonumber \\
&=& \sum_{n'=1}^{L-1}(-1)^{L+n'-1}Y_{jn'}\tA_{L,n'}(k)
 = \sum_{n=1}^{L-1}(-1)^{L+n'-1} Y_{jn}\tA_{L,n}(k)
\eea

We now proceed to prove lemma 2. Using the
decomosition (\ref{det.decomp.k+j}), and
 eq.(\ref{det.k+j-det.k-det.j}) and its similar forms,
 we may write the LHS of lemma 2 as
\bea
&&\left[Y_{jj}\cdot \det_{1,2,\cdots,L-1} +
 \sum_{n=1}^{L-1} (-1)^{L+n}Y_{jn}\tA_{L,n}(j) \right]\cdot \nonumber \\
&& \;\;\;\; \cdot \left[(Y_{kl}+Y_{lk})\cdot\det_{1,2,\cdots,L-1}
 +\sum_{n=1}^{L-1}(-1)^{L+n}\left(
Y_{kn}\tA_{L,n}(l) + Y_{ln}\tA_{L,n}(k)\right)\right] \nonumber \\
&& \;\; -{1\over 2}\left[(Y_{jk}+Y_{kj})\cdot\det_{1,2,\cdots,L-1}
+ \sum_{n=1}^{L-1}(-1)^{L+n}
\left( Y_{kn}\tA_{L,n}(j)+Y_{jn}\tA_{L,n}(k)\right) \right] \cdot \nonumber \\
&& \;\;\;\; \cdot \left[(Y_{jl}+Y_{lj})\cdot\det_{1,2,\cdots,L-1}
 + \sum_{n=1}^{L-1} (-1)^{L+n}
\left( Y_{ln}\tA_{L,n}(j)+Y_{jn}\tA_{L,n}(l)\right)\right] .
\eea
Using eq.(\ref{det.k.j}), we see that the corresponding $G$, similar
 to (\ref{Gmat}), of $\det_{1,2,\cdots,L-1,j,\la k+l \ra}$ has the form
\bea
G_{11} &=& \sum_{n=1}^{L-1} (-1)^{L+n-1} Y_{jn}\tA_{L,n}(j)
 / \det_{1,2,\cdots,L-1} \nonumber \\
G_{12} &=& \sum_{n=1}^{L-1} (-1)^{L+n-1} Y_{jn}[\tA_{L,n}(k)+\tA_{L,n}(l)]
 / \det_{1,2,\cdots,L-1} \nonumber \\
G_{21} &=& \sum_{n=1}^{L-1} (-1)^{L+n-1} [Y_{kn}\tA_{L,n}(j)
+Y_{ln}\tA_{L,n}(j)] / \det_{1,2,\cdots,L-1} \nonumber \\
G_{22} &=& \sum_{n=1}^{L-1} (-1)^{L+n-1} \left(Y_{kn}+Y_{ln}\right)
 [ \tA_{L,n}(k)+\tA_{L,n}(l) ]. \label{Gmat2}
\eea
Using similar expressions, we can write the RHS of  lemma 2 as
\bea
&&\left[Y_{jj}\cdot \det_{1,2,\cdots,L-1}
+\sum_{n=1}^{L-1}(-1)^{L+n}Y_{jn}\tA_{L,n}(j)\right]\cdot \nonumber \\
&& \;\;\;\; \cdot \left[ (Y_{kk}+Y_{lk}+Y_{kl}+Y_{ll})\cdot
\det_{1,2,\cdots,L-1} +\sum_{n=1}^{L-1} (Y_{kn}+Y_{ln})
\left(\tA_{L,n}(k)+\tA_{L,n}(l)\right)\right]\nonumber \\
&& \;\; - \left[(Y_{jk}+Y_{jl})\cdot \det_{1,2,\cdots,L-1}
+ \sum_{n=1}^{L-1}(-1)^{L+n} Y_{jn}\left(\tA_{L,n}(k)
 +\tA_{L,n}(l)\right) \right] \cdot \nonumber \\
&& \;\; \cdot \left[(Y_{kj}+Y_{lj})\cdot \det_{1,2,\cdots,L-1}
 +\sum_{n=1}^{L-1}(-1)^{L+n}\left( Y_{kn}\tA_{L,n}(j)+
 Y_{ln}\tA_{L,n}(j) \right) \right] \nonumber \\
&&  -\left\{ \left[Y_{jj}\cdot\det_{1,2,\cdots,L-1}
 +\sum_{n=1}^{L-1}(-1)^{L+n}Y_{jn}\tA_{L,n}(j) \right]
\left[ Y_{kk}\cdot \det_{1,2,\cdots,L-1}
+\sum_{n=1}^{L-1}(-1)^{L+n} Y_{kn} \tA_{L,n}(k) \right] \right. \nonumber \\
&& \;\;\;\; - \left[Y_{jk} \cdot \det_{1,2,\cdots,L-1}
 +\sum_{n=1}^{L-1}(-1)^{L+n}Y_{jn}\tA_{L,n}(k) \right]
 \left[ Y_{kj} \cdot \det_{1,2,\cdots,L-1}
 +\sum_{n=1}^{L-1}(-1)^{L+n}Y_{kn}\tA_{L,n}(j) \right]\nonumber \\
&& \;\; + \left[Y_{jj}\cdot\det_{1,2,\cdots,L-1}
 +\sum_{n=1}^{L-1}(-1)^{L+n}Y_{jn}\tA_{L,n}(j) \right]
\left[ Y_{ll}\cdot \det_{1,2,\cdots,L-1}
+\sum_{n=1}^{L-1}(-1)^{L+n} Y_{ln} \tA_{L,n}(l) \right] \nonumber \\
&& \left. \;\;\;\; - \left[Y_{jl} \cdot \det_{1,2,\cdots,L-1}
 +\sum_{n=1}^{L-1}(-1)^{L+n}Y_{jn}\tA_{L,n}(l) \right]
 \left[ Y_{lj}\cdot \det_{1,2,\cdots,L-1}
 +\sum_{n=1}^{L-1}(-1)^{L+n}Y_{ln}\tA_{L,n}(j) \right]
\right\}\nonumber \\
&& = \left[Y_{jj}\cdot\det_{1,2,\cdots,L-1} +\sum_{n=1}^{L-1}
(-1)^{L+n}Y_{jn}\tA_{L,n}(j)\right]\cdot\nonumber \\
&& \;\;\;\; \cdot\left[(Y_{kl}+Y_{lk})\cdot\det_{1,2,\cdots,L-1}
+\sum_{n=1}^{L-1}(-1)^{L+n}\left(Y_{kn}\tA_{L,n}(l)+
 Y_{ln}\tA_{L,n}(k)\right)\right]\nonumber \\
&& \; -\left[Y_{jl}\cdot\det_{1,2,\cdots,L-1} +\sum_{n=1}^{L-1}
 (-1)^{L+n}Y_{jn}\tA_{L,n}(l)\right]
\left[Y_{kj}\cdot\det_{1,2,\cdots,L-1} +\sum_{n=1}^{L-1}
(-1)^{L+n} Y_{kn}\tA_{L,n}(j) \right] \nonumber \\
&& \; -\left[Y_{jk}\cdot\det_{1,2,\cdots,L-1} +\sum_{n=1}^{L-1}
 (-1)^{L+n}Y_{jn}\tA_{L,n}(k)\right]
\left[Y_{lj}\cdot\det_{1,2,\cdots,L-1} +\sum_{n=1}^{L-1}
(-1)^{L+n} Y_{ln}\tA_{L,n}(j) \right]
\eea

If we subtract the LHS of lemma 2 by the RHS of lemma 2,  we get
\bea
&& LHS|_{\rm lemma~2} - RHS|_{\rm lemma~2} \nonumber \\
&& \;\;\; = -{1\over 2}\left[ Y_{jk}-Y_{kj}
 +\sum_{n=1}^{L-1}(-1)^{L+n}\left( Y_{jn}\tA_{L,n}(k)-
 Y_{kn}\tA_{L,n}(j) \right) \right] \cdot \nonumber \\
&& \;\;\;\;\; \cdot\left[ Y_{jl}-Y_{lj}
 +\sum_{n=1}^{L-1}(-1)^{L+n}\left( Y_{jn}\tA_{L,n}(l)-
 Y_{ln}\tA_{L,n}(j) \right) \right] \nonumber \\
&& \;\;\; = -{1\over 2}\left[ {\rm det}\left({\b Y}_{L-1}
(k_{L_c},j_{L_r})\right)-
{\rm det}\left({\b Y}_{L-1}(j_{L_c},k_{L_r})\right) \right] \cdot \nonumber \\
&& \;\;\;\;\; \cdot \left[ {\rm det}\left({\b Y}_{L-1}(l_{L_c},j_{L_r})\right)-
 {\rm det}\left({\b Y}_{L-1}(j_{L_c},l_{L_r})\right) \right]
\eea
Again, if the matrix ${\b Y}$ is symmetric, the difference between the LHS and
 RHS of lemma 2 vanishes. This finish the proof of lemma 2.

Finally, for the $L=1$ case (where
$\det_{1,2,\cdots,L-1} = 1$ and $\det_{1,2,\cdots,L-1,j} = Y_{jj}$),
we may verify lemmas 1 and 2 by direct substitutions
 with the abbreviations $\det_{1,2,\cdots,L-1} = \det_0$
 and $\det_{1,2,\cdots, L-1,j} = \det_{0,j} = \det_j$ when $L=1$.

\end{document}